\documentclass[twocolumn,showpacs,preprintnumbers,amsmath,amssymb,a4paper]{revtex4}

\usepackage{graphicx}
\usepackage{dcolumn}
\usepackage{bm}

\newcommand{\bfj}{{\bf j}}
\newcommand{\bfk}{{\bf k}}
\newcommand{\bfq}{{\bf q}}

\begin{document}


\title{Unconventional pairing phases in the two-dimensional attractive Hubbard
model with population imbalance.}

\author{Stefan M.A. Rombouts}
 \email{rombouts@iem.cfmac.csic.es}
\affiliation{%
  Lundbeck Foundation Theoretical Center for Quantum System Research,
  Department of Physics and Astronomy, University of Aarhus,
  DK-8000 Aarhus C, Denmark
}%
\affiliation{%
 Instituto de Estructura de la Materia,
  C.S.I.C.,
  Serrano 123, E-28006 Madrid, Spain}%

\date{\today}

\begin{abstract}
The ground state phase diagram of the two-dimensional attractive Hubbard model
with population imbalance
is explored using a mean field ansatz.
A linear programming algorithm is used to identify the blocked states,
such that the population imbalance can be imposed exactly.
This allows to explore regions of the number-projected phase diagram
that can not be obtained with the conventional Bogoliubov-de Gennes ansatz.
The Fulde-Ferrel-Larkin-Ovchinnikov (FFLO) phase of pairs with non-zero momentum
is found to be the ground state over a wide range of parameters,
while phase separation occurs only in a limited region at small population imbalance.
Through a particle-hole transformation these results can be related
to the underdoped repulsive Hubbard model,
where the FFLO phase takes the form of a particle-hole condensate
that exhibits a spontaneous restoration of spin symmetry.
\end{abstract}

\maketitle


The Bardeen, Cooper and Schrieffer (BCS) theory of
superconductivity \cite{Bardeen57} has been very successful in explaining not only the properties of superconductors
but also many properties of atomic nuclei \cite{Dean03}
and in recent years also properties of ultracold Fermi gases.
%
In this latter respect a particularly interesting setting is offered by optical lattices,
because they allow an unprecedented control over the interaction parameters and geometry of
strongly interacting quantum systems \cite{Kohl05}.
The most important degrees of freedom of such a system are well described by
the minimal Hubbard Hamiltonian \cite{Hubbard63},
\begin{equation}
  H = -  t \sum_{(i,j), \sigma} a^{+}_{i \sigma}  a_{j \sigma}
        + U \sum_i a^{+}_{i \uparrow} a^{+}_{i \downarrow}  a_{i \downarrow} a_{i \uparrow}.
  \label{eq:hubbardmodel}
\end{equation}
The first summation runs over nearest neighbors only.
The Hubbard interaction parameter $U$ reflects the short-range interaction between
particles at the same lattice site.
In the atomic physics setting 
$a^{+}_{i \sigma}$ creates an atom 
in the optical lattice site $i$
and the spin $\sigma$ 
distinguishes between two different types of atoms
or between two distinct hyperfine levels \cite{Lewenstein07}.
In condensed matter physics
$a^{+}_{i \sigma}$ creates an electron with spin-projection $\sigma$
in the Wannier orbital centered at lattice site $i$.
This model is relevant for high-$T_c$ superconductors
where the superconductivity is believed to originate from electrons or holes moving
in two-dimensional cupper-oxide lattices \cite{Anderson92}.

For a system with an equal number of spin-up and spin-down particles
and a negative $U$, i.e. an attractive interaction,
particles of opposite spin pair up to form zero-momentum pairs
that can condense \cite{Paiva04}.
If there is an asymmetry in the particle numbers of both spins,
e.g. an excess of spin-up particles,
then not every spin-up particle will find a spin-down partner to pair up.
It is not yet fully understood what happens in that case.
On the one hand the unpaired particles want to reduce their kinetic energy,
but on the other hand they block part of the phase space
where the system wants to develop pairing correlations.
These conflicting tendencies have inspired many scenarios \cite{Giorgini08}:
the simple BCS picture, where the excess particles minimize their kinetic energy
but block the pairing correlations up to some extent;
breached pairing, where the excess partices are pushed to higher kinetic energies
in order to permit full pairing correlations at the Fermi surface of the minority component
or where some particles of the minority component move to higher kinetic energies
in order to develop pairing correlations at the Fermi surface \cite{Sarma63,Liu03};
or deformed Fermi surfaces, such that minority and majority Fermi surfaces coincide
in certain directions \cite{Muther02}.
Already in the sixties Fulde and Ferrel and independently Larkin and Ovchinnikov
suggested another possibility: non-zero momentum pairs such that pairing correlations
can develop between pairs at different kinetic energies,
and thus bridge the gap between the Fermi surfaces of the two species \cite{Fulde64,Larkin64,Casalbuoni04}.
Even though convincing arguments have been made
that such a state might be energetically favorable \cite{Batrouni08,Bulgac08},
it has been elusive to demonstrate experimentally,
and only recently experiments have provided some evidence for its existence in the
quasi-twodimensional heavy fermion superconductor CeCoIn$_5$ \cite{Kakuyanagi05}.
Another possibility is that the excess particles 
form unpaired regions in real space \cite{Bedaque03,Tempere07}.
The first experiments with unbalanced mixtures of ultracold fermions in traps
exhibited such {\em phase separation} \cite{Zwierlein06,Partridge06}.
Which of the above scenarios prevails will depend sensitively
on the population imbalance, the interaction strength
and the shape of the kinetic energy dispersion curve.
Particularly, it has been shown that the FFLO scenario
is the preferred configuration in the strongly interacting limit \cite{Dukelsky06}.
Here I present a new algorithm to determine the blocked states
at the mean field level.
This method is then applied to the attractive Hubbard model in two dimensions.
Previous studies have already indicated
that in the Hubbard model with imbalanced spin populations
the FFLO phase gains stability near half-filling
because non-zero momentum pairs can maximally exploit
the enhanced phase space offered 
by the nesting of the Fermi surfaces and the corresponding Van Hove singularity
in the middle of the band \cite{Koponen07}.
The ground state phase diagram in this regime
is of particular importance because it might be accessible
in experiments with ultracold atoms in optical lattices
and it can be related to the phase diagram of the repulsive
Hubbard model through a unitary particle-hole transformation,
which might shed light on the pairing mechanism in High-$T_c$ superconductors \cite{Moreo07}.


Mean-field equations can be obtained from a variational ansatz
with a trial state $|\Psi \rangle$ for which Wick's theorem holds,
such that the expectation value of the interaction can be split
into a sum of density $\times$ density terms
$U \langle \Psi | a^{+}_{i \uparrow} a_{i \uparrow} | \Psi \rangle
   \langle \Psi | a^{+}_{i \downarrow} a_{i \downarrow} | \Psi \rangle 
- U \langle \Psi | a^{+}_{i \uparrow} a_{i \downarrow} | \Psi \rangle
   \langle \Psi | a^{+}_{i \downarrow} a_{i \uparrow} | \Psi \rangle
$
(the {\em Hartree terms})
and pairing terms
$U \langle \Psi | a^{+}_{i \uparrow} a^{+}_{i \downarrow} | \Psi \rangle
   \langle \Psi | a_{i \downarrow} a_{i \uparrow} | \Psi \rangle $.
A variation on the trial state leads 
to the Bogoliubov-de Gennes equations,
also known as the Hartree-Fock-Bogoliubov equations.
A restriction of the formalism 
to states that do not mix spin-up and spin-down components
and to the Fulde-Ferrel modulation,
$  \langle \Psi | a^{+}_{j \uparrow} a^{+}_{j \downarrow} | \Psi \rangle
       = e^{i \bfq \cdot \bfj} \Delta_q/U$,
is still general enough 
to compare BCS, breached and FFLO phases and phase separation
and to identify the relevant degrees of freedom.
Minimization of the energy to $ |\Psi \rangle $
for $N_\uparrow$ spin-up and $N_\downarrow$ spin-down particles
in a $L \times L$ lattice,
leads to a mean field Hamiltonian of the form
\begin{eqnarray}
 & & H_{mf,q} =  \sum_{\bfk} \left[ \left( e_\bfk + d_\bfk \right)
                     c^{+}_{\bfk \uparrow} c_{\bfk \uparrow}
        +  \left( e_\bfk - d_\bfk \right)
                     c^{+}_{\bfq - \bfk \downarrow} c_{\bfq - \bfk \downarrow}
  \right. \nonumber \\ & & \left.
        + \Delta_q c^{+}_{\bfk \uparrow} c^{+}_{\bfq -\bfk \downarrow }
        + \Delta_q^{*} c_{\bfq -\bfk \downarrow} c_{\bfk \uparrow} \right]
        + g N_\uparrow N_\downarrow + \frac{\Delta_q^2}{g}
  \label{eq:hmfq}
\end{eqnarray}
with
$g=-U/L^2$,
$ e_\bfk = \left( \epsilon_\bfk + \epsilon_{\bfq-\bfk} 
                  - g (N_{\uparrow} + N_{\downarrow}) \right)/2 - \mu$
and
$ d_\bfk = \left( \epsilon_\bfk - \epsilon_{\bfq-\bfk} 
                  + g (N_{\uparrow} - N_{\downarrow}) \right)/2$.
The eigenproblem of Eq.(\ref{eq:hmfq}) for the unblocked momentum states
reduces to the BCS form,
except that the pairs now have a finite momentum $\bfq$.
The eigenstates are solutions of the BCS-like equations
\begin{equation}
  \sum_{\bfk \ {unblocked}} \frac{1}{E_\bfk} = \frac{2}{g},
\end{equation}
where $E_\bfk^2 = e_\bfk^2 + \Delta_q^2$.
The occupation numbers are 
$  n_{\bfk, \uparrow} = n_{\bfq-\bfk, \downarrow} = \left(1 - e_\bfk / E_\bfk \right)/2 $
for the unblocked momenta,
and 0 or 1 for the blocked momenta.
The Lagrange multiplier $\mu$ has to be adjusted such that 
$\sum_\bfk \left( n_{\bfk, \uparrow} + n_{\bfk, \downarrow} \right)
        = N_\uparrow + N_\downarrow$.
An imbalance in spin populations, e.g. $N_\uparrow > N_\downarrow$,
leads to complications due to the blocking by unpaired particles \cite{Bertsch08}.
Rather than using a different $\mu_\sigma$ for each spin
and fix $N_\uparrow-N_\downarrow$ on average,
here the difference between the particle numbers is imposed exactly
by explicitly selecting a set of blocked momenta.
As shown below this allows to reach certain configurations
that can not be obtained with spin-dependent Lagrange multipliers.
How to select the blocked momenta is a subtle question
because each configuration of unpaired particles
leads to another eigenstate of the mean field Hamiltonian.
For a given value of $\mu$ and $\Delta_q$
the optimal configuration can be found efficiently
using a variation of the Needleman-Wunsch algorithm \cite{Needleman70},
well known for its application to genetic sequence alignment.
To this end one imagines a graph where
the momentum states are ordered on the horizontal axis,
and the sum of $n_\uparrow-n_\downarrow$ for all previous momentum states
appears on the vertical axis.
For each momentum $\bfk$ one has three options:
a pair state with energy $e_\bfk - E_k$,
an unpaired spin-up particle with energy $e_\bfk + d_\bfk$,
and an unpaired spin-down particle with energy $e_\bfk - d_{\bfk}$,
represented respectively by a horizontal jump,
a diagonal jump upwards and a diagonal jump downwards to the right.
There are an exponential number of possible paths
that lead to the desired population imbalance $N_\uparrow-N_\downarrow$,
each with a different total energy.
The optimal path can be split at any momentum $\bfk$:
the segment up to that point on the graph necessarily 
has to be the optimal path to get from the origin to that point;
therefore one has to keep track only of the optimal paths for each point,
and these can be constructed by comparing at most three possible paths 
that arrive in each point of the graph.
In this way the ground state for a given value of $\Delta_q$
can be obtained in ${\cal{O}}\left(3 (N_\uparrow+N_\downarrow) L^2\right)$ operations.
By iterating this procedure for each vector $\bfq$ on the reciprocal lattice,
for several initial values of $\Delta_q$,
one can determine the absolute ground state.
%

\begin{figure}
  \includegraphics[width=8.3cm]{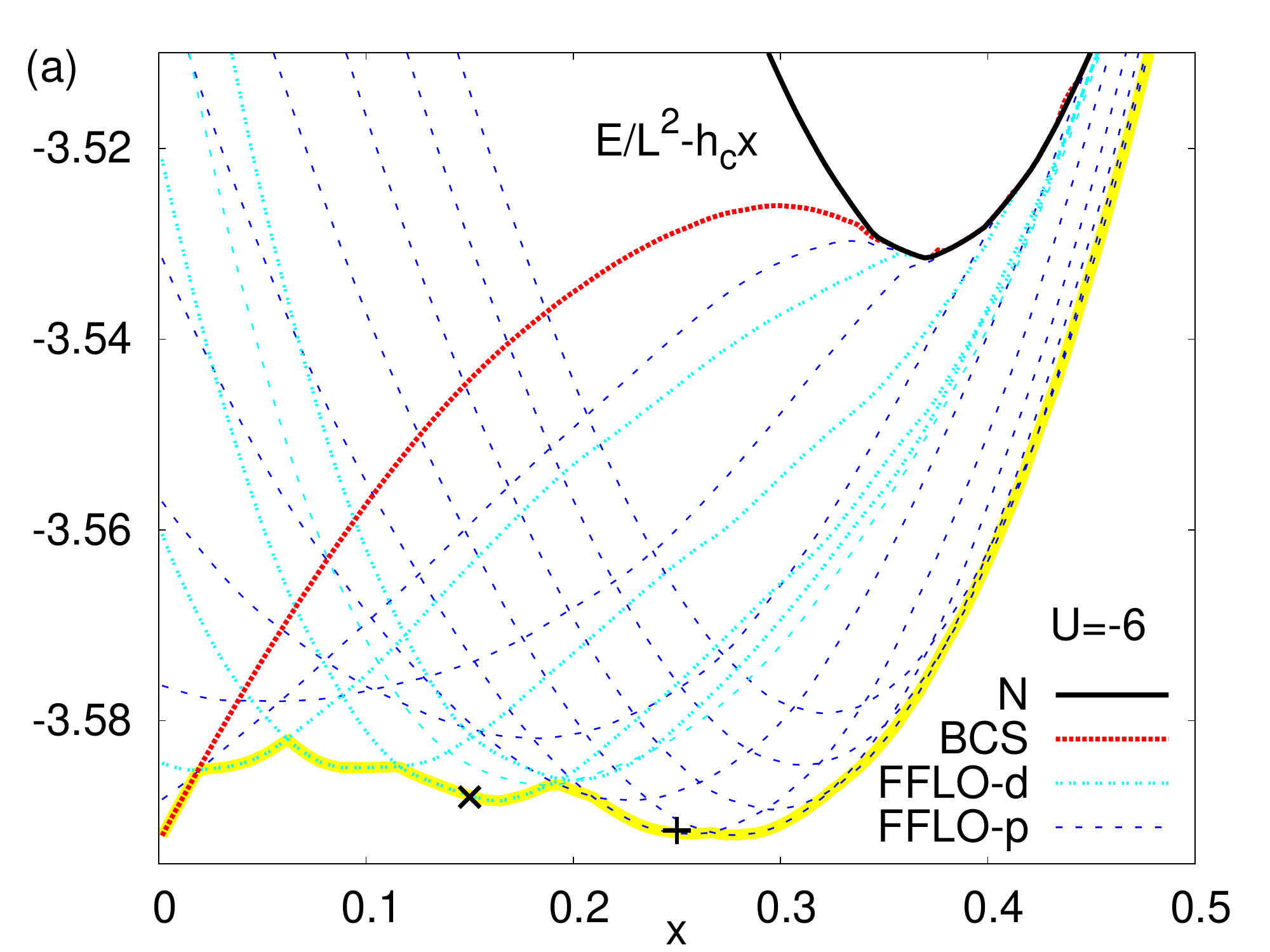}
  \includegraphics[width=8.3cm]{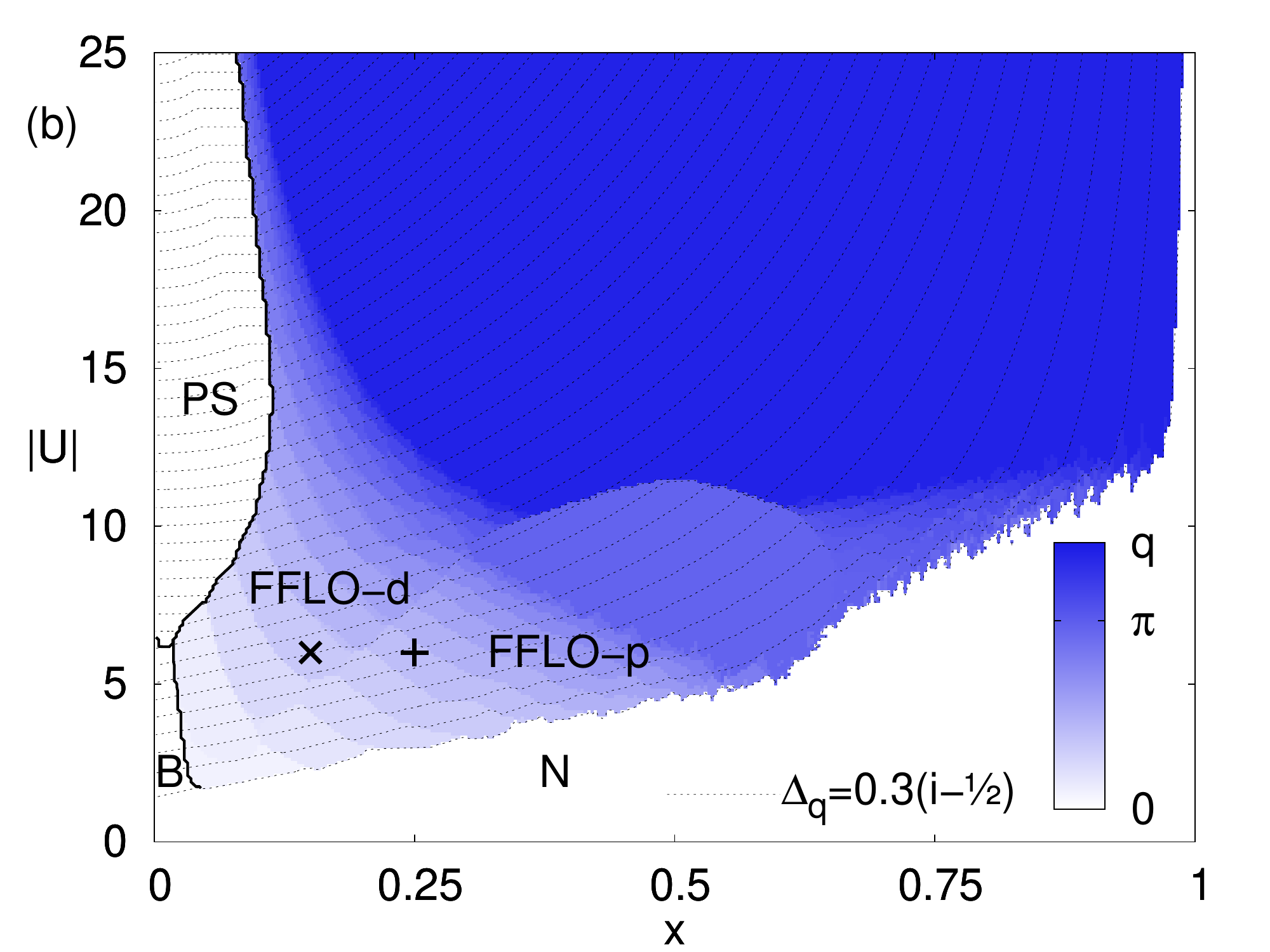}
  \caption{\label{fig:phasediagram} 
      Mean field results for the half filled $25 \times 25$ 
      attractive Hubbard model with population imbalance:
      (a)  Energy per site shifted by $h_c x$,
           for various configurations at $U=-6$;
      (b)  number projected ground state phase diagram.
      Configurations N, B, PS, FFLO-d and FFLO-p explained in the text.
      The momentum densities for the points marked by $\times$ and $+$
      are shown in Fig.(\ref{fig:densities}).
  }
\end{figure}

The zero temperature phase diagram
of the half-filled 2D attractive Hubbard model 
was determined by evaluating the ground state for a $25 \times 25$ lattice
for $N_\uparrow + N_\downarrow = L^2$, all possible asymmetries 
$x=(N_\uparrow - N_\downarrow)/(N_\uparrow + N_\downarrow)$
and for all possible values of $\bfq$.
Fig.\ref{fig:phasediagram}a shows the evolution of the energy for $U=-6$.
The system starts out in a BCS configuration with zero-momentum pairs (B).
At $x=0.02$ the FFLO configurations with $q>0$ start to take over:
first configurations where $q$ aligns along the diagonal (FFLO-d),
and then at $x=0.19$ configurations with $q$ parallel to a lattice axis (FFLO-p).
At large asymmetries $\Delta_q$ goes to zero
and everything converges to the normal state (N).
For $\bfq>0$ this point is shifted to larger asymmetries.
This means that FFLO mechanism allows the system to
avoid the the Clogston-Chandrasekhar limit \cite{Clogston62,Chandrasekhar62}.
Instead, the system undergoes a crossover 
from the FFLO region to the normal state at a larger asymmetry, 
that depends on the interaction strength $U$.
The minimal energy curve in Fig.\ref{fig:phasediagram}a 
displays a concave behavior for $x<0.3$.
A grand canonical approach based on a difference in Lagrange multipliers,
$h = (\mu_\uparrow - \mu_\downarrow)/2$,
would result in first order transition at $h_c=2.077$
from the balanced BCS configuration to the FFLO-p minimum at $x_c=0.28$,
and would never produce a ground state with particle numbers
corresponding to an asymmetry $x$ in between $0$ and $x_c$.
This means that the results in this range can only be obtained
by imposing the exact particle number difference.
The FFLO states in this region could be stabilized
by long range spatial modulations,
as explained already by Larkin and Ovchinnikov \cite{Larkin64},
and by the residual interaction.
Apart from the N, B, FFLO-d and FFLO-p regions,
the ground state phase diagram in Fig.\ref{fig:phasediagram}b
also shows a region with phase separation (PS),
where the system splits into balanced BCS domains
and fully polarized domains.
The corresponding energy was evaluated as
$ E_{ps} = \upsilon E_f\left(|N_\uparrow-N_\downarrow|/\upsilon \right)
      + (1-\upsilon) E_p\left(N_\downarrow/(1-\upsilon)\right)$,
with $E_f(n)$ the energy of $n$ unpaired spin-up particles
in the $L \times L$ lattice without paired particles
and with $E_p(n)$ the energy of $n$ pairs in the same lattice 
but without unpaired particles,
where $\upsilon$ was optimized to minimize $E_{ps}$.
Note how the value of $\bfq$ evolves as a function of the Hubbard interaction $U$:
$\bfq$ is not a constant function of the difference between the two Fermi momenta,
as suggested by previous mean field studies \cite{Moreo07},
but rather depends on $U$ times $x$ as shown by the bands in Fig.\ref{fig:phasediagram}b.

\begin{figure}
  \includegraphics[width=8.3cm]{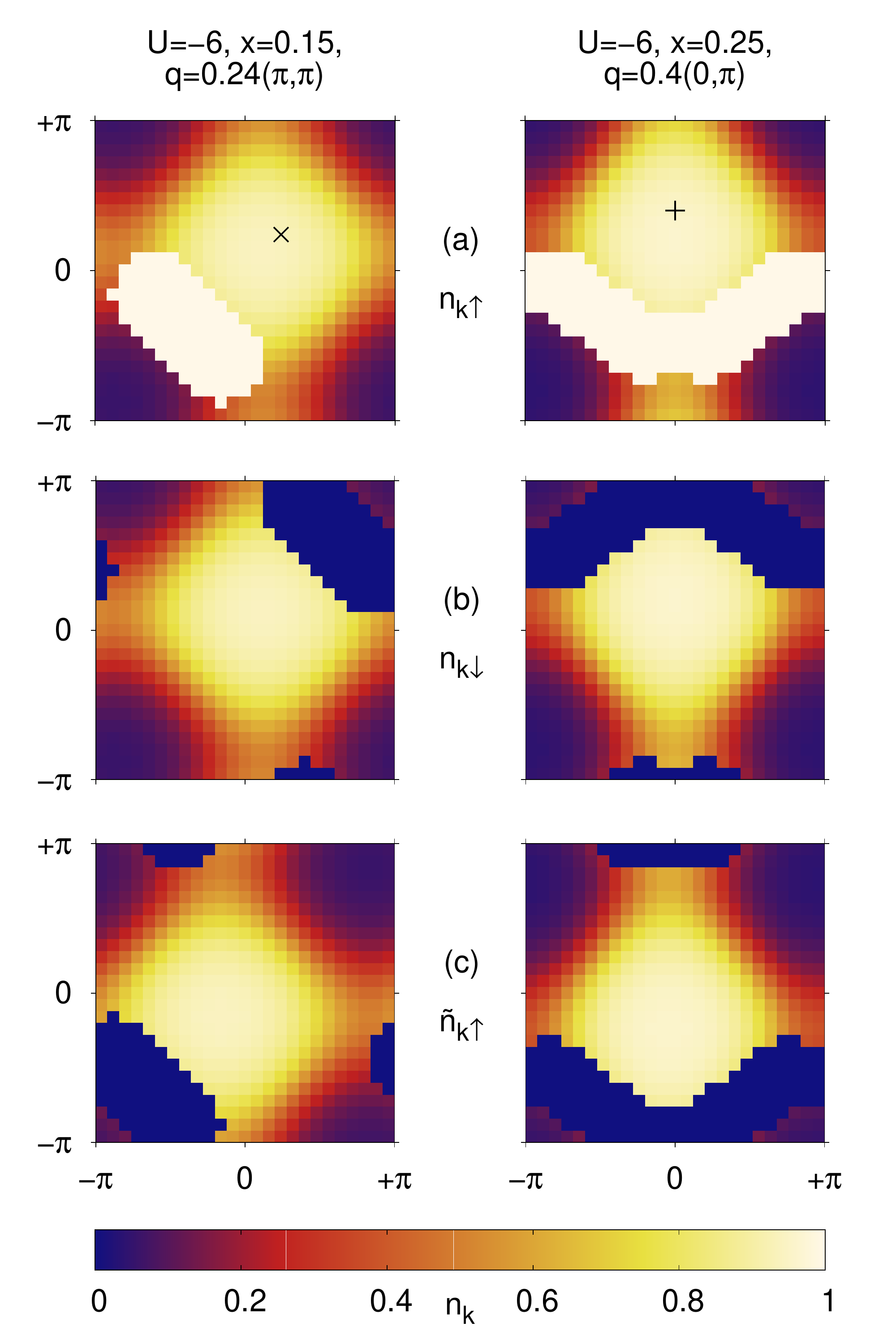}
  \caption{\label{fig:densities} Momentum density profiles
           for the diagonal and parallel FFLO configurations
           marked in Fig.(\ref{fig:phasediagram}):
           (a) spin-up particles,
           (b) spin-down particles
           (c) transformed spin-up holes.
  }
\end{figure}

Momentum density profiles typical for the FFLO-d and FFLO-p regions
are shown in Fig.\ref{fig:densities}.
They clearly show the regions of blocked momenta.
Therefore an obvious way to confirm the FFLO phases experimentally
would be to look for such distinctive asymmetries
in the spin projected momentum densities.
Another option would be to look for signals
in radiofrequency spectroscopy
or momentum resolved photoemission spectroscopy \cite{Bakhtiari08},
although a more advanced many-body treatment might
be necessary in order to identify the FFLO phases accurately.
An interesting point about the FFLO phase is
that the lowest energy degrees of freedom are not given by
Bogoliubov quasiparticles (they are inhibited by the gap $\Delta_q$),
but rather by excitations of the unpaired particles,
for which there is no gap.
The dynamics of these unpaired particles 
is governed by an effective singe-particle potential given by $d_\bfk-E_\bfk$
and a $p$ wave interaction of first order in $U$
induced by the residual part of the full Hamiltonian
through the exchange of spin-down particles with the pair condensate.
The unpaired particles also experience an interaction
of second order in $U$,
induced by phonons in the pair condensate \cite{Bulgac06}.


Finally, I want to draw attention to the implications of these results for
the repulsive Hubbard model:
one can define a particle-hole transformation,
$ \tilde{c}_{\bfk \uparrow} = c^+_{(\pi,\pi)-\bfk \uparrow}$,
$ \tilde{c}_{\bfk \downarrow} = c_{\bfk \downarrow}$,
\cite{Hirsch85},
that converts the half-filled attractive model with population imbalance
into the underdoped repulsive model with equal spin populations.
If the above calculations give a qualitative picture
of the ground state for the imbalanced attractive model,
then the particle-hole transformed equivalent should be relevant
for the ground state of the underdoped repulsive model.
The third panel of Fig.(\ref{fig:densities}) shows the transformed momentum profile
for the spin-up particles.
There is a remarkable symmetry with the momentum profile for the spin-down particles,
that seems to be a general feature of the FFLO phase,
and that makes sense because there is no a priori asymmetry in the repulsive case.
The FFLO pair condensate 
transforms into an asymmetric particle-hole condensate.
The onset of this condensation occurs in the same parameter range
as the {\em pseudogap} \cite{Norman05}
in the underdoped repulsive model,
with an anisotropic gap of the order of the Hubbard interaction.
The residual interaction
could then produce strong hole-hole correlation effects:
the anisotropy of the repulsion between the holes 
would enhance the formation of striped patterns,
in line with the suggestions from Moreo {\em et al.}
on the relation between the attractive FFLO phase
and a {\em striped} phase in the repulsive model \cite{Moreo07}.


Through explicit selection of the blocked momenta,
the particle number difference $N_\uparrow-N_\downarrow$ 
was imposed exactly in the mean field solution.
This has allowed the exploration of
normal, phase separated, BCS pairing, breached pairing and
non-zero momentum Fulde-Ferrel pairing phases 
in the ground state phase diagram
for the two-dimensional attractive Hubbard model
with population imbalance at half filling.
At very small population imbalances,
the ground state evolves from normal to BCS pairing to phase separation
as $|U|$ is increased.
At larger imbalances the phase diagram at intermediate interaction strengths
is dominated by non-zero momentum FFLO pairs,
with a pair momentum $\bfq$ that scales proportionally to the population imbalance
and the interaction strength.
These configurations might be identified experimentally
from asymmetries in the spin projected momentum densities.
Through a particle-hole transformation these results can be related
to the repulsive Hubbard model with equal spin populations below half filling.
There the FFLO mechanism gives rise to the formation of a particle-hole condensate
on which the residual interaction builds strong hole-hole correlations.
This might be related to a pseudogap phase and stripe formation.

I thank K. M\o{}lmer, N. Nygaard, R. Molina and J. Dukelsky
for the interesting discussions.


\end{document}